\documentclass[twocolumn,aps,floatfix,showkeys]{revtex4}

\usepackage{graphicx} 
\usepackage{float}
\usepackage{color}
\usepackage{amssymb}
\newcommand{\AVE}[1]{\left\langle{#1}\right\rangle}

\begin{document}

\title{Resonant impurity states in chemically disordered\\ half-Heusler
  Dirac semimetals}

\author{J.~Kiss}
\affiliation{Max-Planck-Institut f\"ur Chemische Physik fester Stoffe,
  N\"othnitzer Str.~40, 
 01187  Dresden, Germany}
\author{S.~W.~D'Souza}
\affiliation{Max-Planck-Institut f\"ur Chemische Physik fester Stoffe,   N\"othnitzer Str.~40, 
 01187  Dresden, Germany}
\author{L.~Wollmann}
\affiliation{Max-Planck-Institut f\"ur Chemische Physik fester Stoffe,   N\"othnitzer Str.~40, 
 01187  Dresden, Germany}
\author{C.~Felser}
\affiliation{Max-Planck-Institut f\"ur Chemische Physik fester Stoffe,    N\"othnitzer Str.~40, 
 01187  Dresden, Germany}
\author{S.~Chadov}
\affiliation{Max-Planck-Institut f\"ur Chemische Physik fester Stoffe,   N\"othnitzer Str.~40, 
 01187  Dresden, Germany}
\author{K.~Chadova}
\affiliation{Dept. Chemie, Ludwig-Maximilians-Universit\"at, Butenandtstr.~11, 81377 M\"unchen, Germany}
\author{D.~K\"odderitzsch}
\affiliation{Dept. Chemie, Ludwig-Maximilians-Universit\"at,
  Butenandtstr.~11, 81377 M\"unchen, Germany}
\author{J.~Min\'ar}
\affiliation{Dept. Chemie, Ludwig-Maximilians-Universit\"at,
  Butenandtstr.~11, 81377 M\"unchen, Germany}
\author{H.~Ebert}
\affiliation{Dept. Chemie, Ludwig-Maximilians-Universit\"at,
  Butenandtstr.~11, 81377 M\"unchen, Germany}


\begin{abstract}
We  address the electron transport characteristics in bulk half-Heusler
alloys with their compositions tuned 
to the borderline between topologically nontrivial semi-metallic and trivial semiconducting
phases. The precise first-principles calculations based on the
coherent potential approximation (CPA) reveal that all the studied
systems exhibit sets of dispersionless impurity-like resonant
levels, with one of them being located right at the Dirac point. By means of the Kubo formalism we
reveal that the residual conductivity of these alloys is strongly
suppressed by impurity scattering, whereas the spin Hall
conductivity exhibits a large value which is comparable to that of Pt,
thereby leading to divergent spin Hall angles.
\end{abstract}

\maketitle

\section{Introduction}

The subgroup of the half-Heusler materials with a ternary composition 
(consisting of an alkali- or an early transition element, main-group
element and a late transition element) contains a large family of 
nonmagnetic semiconductors and nonmagnetic semimetals. Exceptions are the
half-Heuslers containing Mn as an early transition element (which always
tends to develop a strong localized magnetic moment), or those, whose number of valence electrons differs from 18 per
formula unit. The variety of possible ternary atomic combinations makes
the half-Heusler family large enough to include topologically distinct classes of materials - trivial (typically light compounds exhibiting a direct band gap at the
$\Gamma$ point of the Brillouin zone) and nontrivial systems  (semimetals or gapless semiconductors
exhibiting the band inversion at the $\Gamma$ point) with respect to the
$\mathbb{Z}_2$ invariant, which have been studied in details in Refs.~\cite{CQK+10,LWX+10,SLM+10}.
On the other hand, those systems which might lie
at the borderline between the nontrivial and trivial systems,
cannot be strictly identified, despite that they may exist within this group. The reason
for this is quite clear: in contrast to e.g., Weyl semimetals, for the
half-Heusler and the related zinc-blende structure types there are no strict symmetry arguments  which would guarantee the stability of the topological
transition state. Because of that, the emergence of linear-dispersive states which
might manifest the topological phase transition in the half-Heusler
systems can be expected only at the nontrivial/trivial interfaces. 
Nevertheless, the borderline bulk systems in principle can be constructed within the
Heusler class, since there exist several physically
reasonable ``adiabatic paths'' which continuously link both topological
classes, such as pressure and chemical substitution. Some of such systems,
as e.g., the zinc-blende Hg$_{0.83}$Cd$_{0.17}$Te, where already quite
clearly identified in the past~\cite{TT84,THS85} and also studied recently~\cite{OBZ+14}. Their
Dirac cones centered at the $\Gamma$ point of the Brillouin zone represent a separate class of massless fermions (Kane fermions) which cannot
be reduced to any well-known case of massless particles in quantum
electrodynamics~\cite{OBZ+14}. In addition, since the Heusler-based topological
systems attract growing interest due to the sensitivity of their
properties and simultaneous stability of the structure with respect to
the chemical content, we will model the chemical variety
using  first-principles calculations by bringing the system into
the topological borderline regime to study its residual transport characteristics.

\section{Modeling}

After first theoretical attempts to classify  the half-Heusler semiconductors in
the sense of the topological $\mathbb{Z}_2$ invariant~\cite{CQK+10,LWX+10}, many similar calculations have been
performed. Since the numerical methods as well as the ways in which they have been applied are very
different, the results for a particular half-Heusler composition may vary a
lot. On the other hand, all these results exhibit the same general trend: the light half-Heuslers appear to be trivial, whereas
the heavy ones are most probably nontrivial. In this context, there
is no guiding principle used to select  particular systems we are going to
link by an appropriate chemical substitution, but we based our modeling using the systems which are already known.
The first system is the heavy half-Heusler LaPtBi~\cite{HSR+02}, which in most
of the calculations exhibits a band-inversion~\cite{Ogu01} indicating that it
represents a typical nontrivial semimetal. The lighter half-Heusler
LaPdBi~\cite{SKMY07}, which in calculations exhibits a small indirect band gap~\cite{CQK+10,LWX+10,SLM+10},
will serve as a representative trivial system. Since both compounds have the same
 crystal structure, it might be possible to generate a series of isostructural
 La(Pt$_{1-x}$Pd$_{x}$)Bi alloys, by continuously varying $x$ from 0 to 1. Such series will necessarily contain
a specific composition corresponding to the topological
transition. Theoretically we can efficiently search for it by performing
a series of calculations for different concentrations and inspecting the change of the electronic
structure. Similar calculations were performed already for the binary
systems, as e.g. for  the similar class of the zinc-blende~\cite{CKKF12} and the rock-salt
semiconductors~\cite{NSPS14,SBB+15}. 

\section{Calculations}

We performed our calculations using the relativistic Green function SPR-KKR
method~\cite{EKM11} in combination with the CPA alloy theory~\cite{Sov67,Tay67} to study the
nonstoichiometric compositions, and applying the spherical
approximation (ASA) for the atomic potentials. The exchange-correlation
functional was treated within the generalized gradient 
approximation (GGA)~\cite{PBE96}.
The lattice constant as a function of $x$ was interpolated linearly
between the experimental values for the concentrations ${x=0}$ and
1. Since the disorder breaks translational symmetry,
the electronic eigenstates  are not localized in the momentum space. In this case, in order to visualize the electronic
structure we will use the concept of the Bloch spectral function (BSF), which provides  a
distribution of the electron spectral weight within energy-momentum
space, being defined as a Fourier transform of the real-space Green
function $G(\vec r,\vec r\,',E)$: 
\begin{eqnarray*}
A(\vec k,E)&=&-\frac{1}{\pi N}~{\rm Im}\!\!\sum_{n,m=1}^N\!\!e^{i\vec k(\vec R_n - \vec
    R_m)}\\ &\times&\int d^3r \left\langle G(\vec r+\vec R_n,\vec r+\vec R_m, E)\right\rangle\,,
\end{eqnarray*}
where $\left\langle{}\right\rangle$ denotes the CPA average and  $\vec
R_{n(m)}$ are the atomic lattice vectors. $A(\vec k,E)$
can be interpreted as a $\vec k$-resolved DOS (density of states) function, since
\begin{eqnarray*}
n(E) = \frac{1}{\Omega_{\rm BZ}}\int\limits_{\Omega_{\rm BZ}}d^3k\,A(\vec k,E)\,,
\end{eqnarray*}
with $n(E)$ being the total DOS, and $\Omega_{\rm BZ}$ the Brillouin zone volume. 
For the compound with ideal translational symmetry, the Green function contains poles on the
energy axis, which results into a set of the  $\delta$-function peaks
for the DOS, BSF, or other energy-dependent quantities. To keep them for
pure systems as continuous energy
functions, they are calculated via the Green function which is
slightly misplaced from its poles by adding a small
imaginary offset Im$E$\,=\,$\delta$ to the energy argument. If disorder treated by the CPA,
the effective Green function does not contain poles on the real
energy axis due to the effective self-energy, which strictly speaking contains a non-zero
imaginary part at any energy value by making the effective Green function
to be analytical everywhere on the real energy axis. Nevertheless, since here we are
dealing with semimetals, their spectral weight close to
the Fermi energy is very small, which leads to a
severe reduction of the imaginary part of the corresponding
self-energy. To avoid numerical instabilities which might be
possible in such a regime for the BSF calculation we add to the energy a very small imaginary offset
${\delta=0.01}$~meV.

\section{Electronic structure}  

\begin{figure}
\centering
\includegraphics[width=1.0\linewidth]{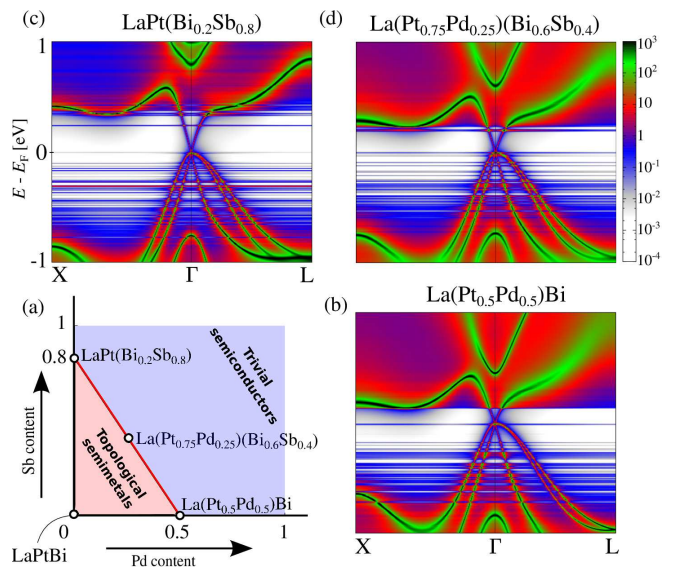}
\caption{(a)~Compositional tuning. By taking LaPtBi 
as a starting point, we generate the series of La(Pt$_{1-x}$Pd$_x$)Bi
alloys, in which the topological phase transition occurs at
${x\approx0.5}$. Analogously, within
LaPt(Bi$_{1-y}$Sb$_y$) series, the transition occurs at ${y\approx0.8}$. These two compositions
generate a continuous set of borderline materials 
La(Pt$_{1-x}$Pd$_x$)(Bi$_{1-y}$Sb$_y$): ${y\approx0.8-1.6\cdot x}$, whose
compositions are placed along the red line separating the nontrivial and trivial systems (light-red and light-blue colored areas, respectively). (b), (c) and (d)~represent the BSFs,
calculated for three borderline compositions: La(Pt$_{0.5}$Pd$_{0.5}$)Bi, LaPt(Bi$_{0.2}$Sb$_{0.8}$) and
La(Pt$_{0.75}$Pd$_{0.25}$)(Bi$_{0.6}$Sb$_{0.4}$), respectively. The
spectral intensity distribution corresponds to the logarithmic color
scale, shown at the right side of (d).}
\label{fig1}
\end{figure}

Schematically the search for the borderline compositions is shown in
Fig.~\ref{fig1}\,a. 
We start with LaPtBi which belongs to the nontrivial phase (point at the origin) and continuously substitute Pt by Pd 
going towards LaPdBi (${x=1}$). The transition into
the trivial phase occurs at about ${x=0.5}$, i.e. for
La(Pt$_{0.5}$Pd$_{0.5}$)Bi.  The corresponding BSF is
depicted in Fig.~\ref{fig1}\,b. Just to demonstrate the flexibility and
variability of the ternary half-Heusler systems, we can select another chemical link
from LaPdBi into the trivial phase, by substituting the main-group
element. For instance, we can continuously substitute Bi by Sb
generating another set of Heusler series, LaPt(Bi$_{1-y}$Sb$_{y}$), which crosses
the topological transition point at about ${y=0.8}$; the corresponding
BSF is depicted in Fig.~\ref{fig1}~c. We notice,
that the stoichiometric compound LaPdSb (${y=1}$) exists in reality
as well, however it crystallizes in the hexagonal structure~\cite{WY12}. To
establish the corresponding series, we used the cubic half-Heusler structure
for LaPdSb with the lattice constant theoretically estimated in Ref.~\cite{SLM+10}.
Two isostructural and isoelectronic borderline compounds,
La(Pt$_{0.5}$Pd$_{0.5}$)Bi and LaPt(Bi$_{0.2}$Sb$_{0.8}$), generate the
 continuous set of La(Pt$_{1-x}$Pd$_{x}$)(Bi$_{1-y}$Sb$_{y}$)
alloys which could be expected to appear at the topological
borderline, if their compositions satisfy  ${y\approx 0.8-1.6\cdot x}$,
marked by the straight red line in Fig.~\ref{fig1}\,a. To ensure this, we 
pick up the composition from the middle of the red line,
i.e. La(Pt$_{0.75}$Pd$_{0.25}$)(Bi$_{0.6}$Sb$_{0.4}$), which exhibits
the close proximity to the phase transition as indicated by its
calculated BSF in Fig.~\ref{fig1}\,d. In principle, it is 
possible to establish one more path from LaPtBi into the trivial phase,
by manipulating the early transition element, i.e. by substituting La
with e.g. Sc. Thus, generally, the manifold of possible borderline
compositions will represent a two-dimensional surface in a three-dimensional parameter space.

By comparing all three BSFs presented in Fig.~\ref{fig1}, one finds several
similar features. One of them is the presence of a Dirac cone centered
almost at the Fermi energy ($E_{\rm F}$) in the $\Gamma$ point of the
Brillouin zone, which indicates their topological phase transition state. Since all compounds show
a considerable amount of chemical disorder, one observes noticeable broadening of
the spectral intensity, especially within the energy regime where the
spectral intensity is strong, i.e. approximately from 0.25~eV above  and
from $-0.8$~eV below $E_{\rm F}$. By approaching $E_{\rm F}$ the  broadening reduces leaving the conical bands to remain
Bloch-like. Another interesting feature are the dispersionless
impurity-like states, recognized as horizontal stripes which become more
distinct within the regime closer to $E_{\rm F}$. They remind 
the resonant impurity levels observed in the gapless semiconductors
as e.g., in PbTe host doped by Tl or Ti, and play an important
role in tuning of thermoelectric properties~\cite{HWC+12}. On a first glance, the 
 cases considered here may seem to be different, since the chemical disorder
rates are very high, thus in the context of the chemical composition the
impurities are formally absent. One can however straightforwardly see
that the impurity-like effects occur at arbitrary disorder
rate, since any chemically-disordered system can be approximated by
the translational-invariant regular system to any limit. This can be
achieved, for instance, by subsequent accumulation of different chemical configurations
within the correspondingly expanded translational supercell.  Once the
difference between the disordered original and the translational-invariant
systems which actually contains randomness (the so-called self-energy,
if the ``difference'' is treated in terms of the Green functions) appears to be small enough,
it must produce nothing else but the effect of random impurities embedded
in the ``host'' represented by the translational-invariant system. 

The  BSFs presented in Figure\,\ref{fig1} reveal, that by moving away from $E_{\rm F}$ the
density of resonant levels increases; in the energy regime where the DOS becomes sufficiently large, they approach
each other close enough to merge into a continuum. This situation
is typical for the conventional metallic alloys, where the distinct resonant
levels cannot be easily observed. In contrast to conventional metals, here 
 in the vicinity of $E_{\rm F}$ the DOS is strongly reduced, and the
 impurity levels become evident. At the same time, they remain resonant (i.e., not isolated) since the semimetals do not exhibit the energy gap. 

\begin{figure}
\centering
\includegraphics[width=1.0\linewidth]{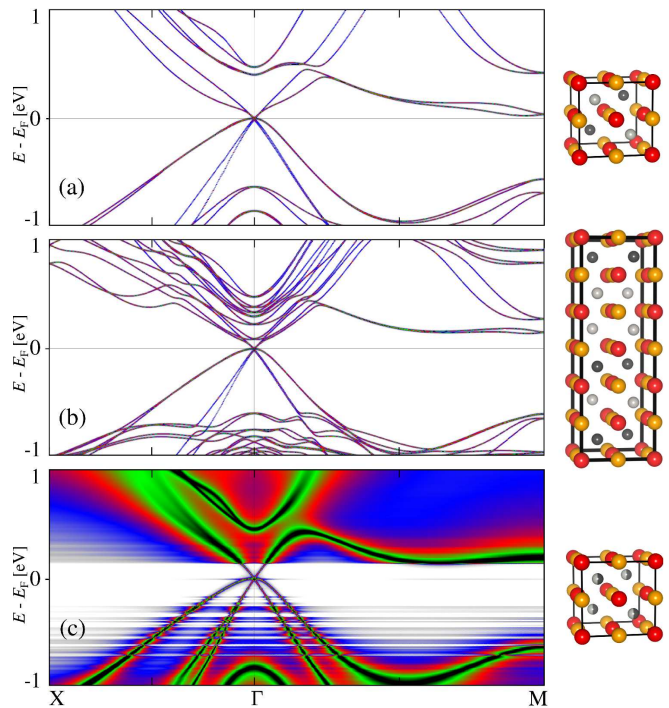}
\caption{Comparison of the electronic structures of the same
  La(Pt$_{0.5}$Pd$_{0.5}$)Bi composition, but related to the different
  chemical orders. Corresponding crystal structure units are shown on the
  left. Red and yellow spheres mark La and Bi atoms, gray and dark-gray - Pd and Pt, respectively. (a)~The smallest possible
  ordered structure (8 nonequivalent atoms within unit cell) of the given stoichiometry, in which half of the
   tetrahedral sites are occupied by Pt, half - by Pd.  (b)~More
   complicated ordered variant of the same composition with 24
   nonequivalent atoms which includes more environmental combinations.
 (c) The system with fully random Pd/Pt disorder (random occupation of
   the tetrahedral sites by Pt/Pd is indicated by the  half-dark/half-light-gray spheres, respectively).}
\label{fig3}
\end{figure}

Based on the same arguments as we have used above, we can decompose our
original disordered system (represented by CPA in Fig.~\,\ref{fig1}\,b
or in Fig.~\,\ref{fig3}\,c) into a translational-invariant ``host'', which is close enough in the sense of 
 the same borderline behavior, plus the  remaining self-energy
 representing randomness. In case of the disordered La(Pt$_{0.5}$Pd$_{0.5}$)Bi alloy, such a ``host''
can be already represented by the minimal ordered system shown in Fig.~\ref{fig3}\,a. We can continue this procedure by including
more random combinations (Fig.~\ref{fig3}\,b) in the unit cell by approaching the
electronic structure to the original situation in Fig.~\ref{fig3}\,c.
The randomness, which in this case becomes relatively small, will have a strong broadening effect 
only within those energy regimes where the large spectral weight of the
host states is contained (at those energy levels, where  many
distinct eigenstates are situated closely to one another), i.e. far away from $E_{\rm F}$, and thus, 
far from the conical bands.  In this context, it becomes clear why these
conical bands are so ``insensitive'' to disorder in terms of CPA. The reason is that
their presence is an intrinsic property of the ordered ``host'' system  and the randomness,
which strongly affects only the energy range with a large DOS, cannot destroy
it. At most, what may occur in the low DOS region, are rather weak disorder
effects, as e.g., the appearance of the distinct impurity levels. 
In particular, one impurity level will be formed in the $\Gamma$ point
at $E_{\rm F}$, due to the four-fold degeneracy of the
corresponding eigenstates.

\begin{figure*}
\centering
\includegraphics[width=1.0\linewidth]{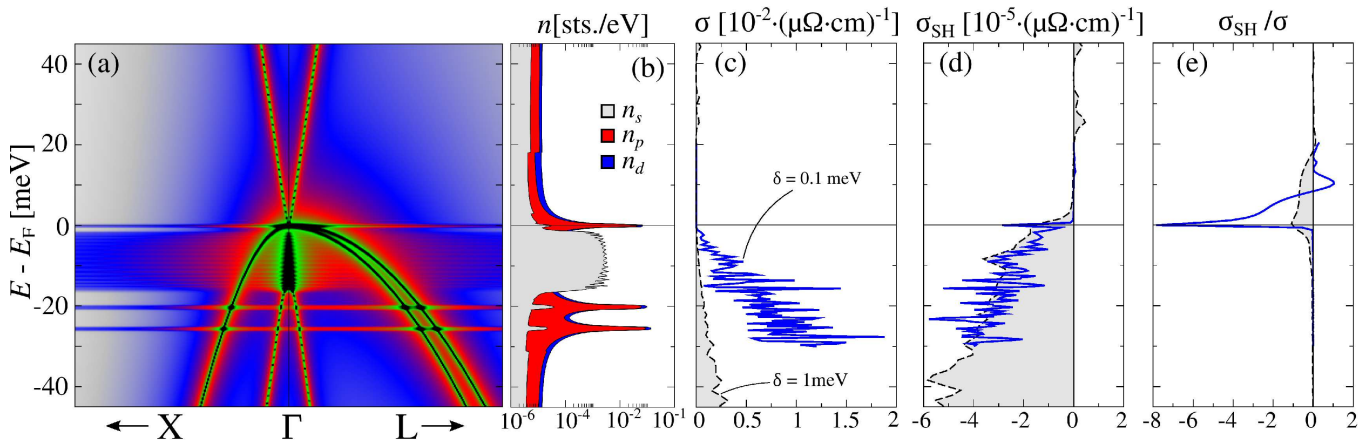}
\caption{(a)~Calculated BSF for La(Pt$_{0.5}$Pd$_{0.5}$)Bi alloy (spectral intensity distribution is the same as in Fig.~\ref{fig1}) and (b)~the corresponding $s$-, $p$- and $d$-projected DOS
  shown using the logarithmic scale ($n_{s,p,d}$ - in gray, red and
  blue, respectively). (c)~Diagonal conductivity $\sigma$,  spin Hall
  conductivity $\sigma_{\rm SH}$, and their ratio $\sigma_{\rm SH}/\sigma$ (spin Hall angle), calculated as functions of the energy $E$. Black dashed line corresponds  to the small imaginary energy offset  ${\delta=1}$~meV, thick blue line -- to ${\delta=0.1}$~meV.}
\label{fig2}
\end{figure*}

By inspecting the BSF in the meV regime (Figs.~\ref{fig2}\,a, b)
 the structure of the impurity level in the vicinity of  $E_{\rm F}$  appears to be  rather complicated: it consists of three distinct peaks
produced mostly by mixture of the $p_{3/2}$- and the $d$-states as well as of the relatively broad set of smaller peaks of the $s$-states spread
almost equally over all atomic types. It indicates that the impurity levels
 stem not just from particular atoms, but rather from hybridized  orbitals which include contributions from all types of atoms constituting the formula unit, as it
 occurs in the zinc-blende, rock-salt and thus in the Heusler structures.

We notice, that all above considerations are restricted only to  those
types of disorder, which affect the translational symmetry, but not
the chemistry of a compound or the topology of a lattice (so, that a
system can still  be considered as an 18 electron half-Heusler). 

On the next step we will study the transport characteristics, which appear to be unusual due to 
a rich combination of different electronic structure features in the vicinity of $E_{\rm F}$. 

\section{Transport characteristics}

Here we will study the electron transport characteristics of
La(Pt$_{0.5}$Pd$_{0.5}$)Bi by calculating the residual conductivity by means of the general Kubo 
formalism~\cite{But85,LGK+11}: 
\begin{eqnarray}
  \sigma_{\mu\nu} &=& \frac{\hbar}{\pi V}{\rm Tr}\AVE{\hat{J}_{\mu}\Im
    G^+\,\hat{\!j}_\nu\Im G^+}\nonumber \\ & + & \frac{i\hbar}{2\pi V}{\rm Tr}\AVE{\left[
      \hat{J}_{\mu}\Im G^+\,\hat{\!j}_\nu- \hat{J}_{\nu}\Im G^+\,\hat{\!j}_\mu
      \right]\Re G^+}\nonumber \\
    &+& \frac{e}{2\pi V}{\rm Tr}\AVE{\Im G^+\left(r_{\mu}\hat{J}_{\nu} -
     r_{\nu}\hat{J}_{\mu}\right)}\,,
\label{eq:Kubo-Streda-1}
\end{eqnarray}
where $\Re G^+$ and  $\Im G^+$ are the Hermitian and anti-Hermitian components
of the retarded Green function $G^+$; $\hat{J_{\mu}}$ and
$\,\hat{\!j}_{\mu}$ are the spatial $\mu$-components of the general
response, i.e.  spin- and
charge-currents, respectively. The brackets $\AVE{}$ mean the
configurational average required in case of random disorder.  
The first term (so-called Kubo-Greenwood term) is symmetric and thus is responsible for the longitudinal
conductivity (expressed by the diagonal part of the conductivity
tensor). The second and the third terms are antisymmetric and thus contribute to 
the off-diagonal conductivities. Since the first and the second terms contain the configurational average over the
products of Green functions, in case of disorder they normally need systematic
improvement (the so-called vertex corrections, which we take into
account). This is, however is not necessary for the last term - its configurational average reduces to
the average of single Green function as delivered by CPA.
Assuming the zero-temperature case, the longitudinal conductivity is precisely given by the first term evaluated
at $E_{\rm F}$. In case of the zero DOS at $E_{\rm F}$ (i.e., in the
semiconductor case) it is precisely zero. The same holds for the second
(antisymmetric) term. The third term in general is non-zero, even in case
of the zero-temperature semiconductor, as it contains contributions from the occupied
states (so-called ``Fermi-sea'' contribution). It was demonstrated by
St\v{r}eda~\cite{Str82} that it is possible to include the Fermi-sea
contributions by using the Green function at $E_{\rm F}$ due to its
analytical properties. The St\v{r}eda formulation is
exact only for the conserving currents (the transverse charge currents,
such as e.g. the Fermi-sea contributions to the ordinary or anomalous
Hall effects). For the spin current, which in general is a non-conserving
quantity in the bulk, it is only an approximation which assumes that the
non-conserving component of the spin current vanishes.

In order to access the spin-resolved off-diagonal components $\sigma_{xy}^{\uparrow(\downarrow)}$, where the spin polarization $\uparrow(\downarrow)$ refers
to the $z$-axis, we employed the relativistic scheme~\cite{VGW07,LGK+11}.
Due to the small DOS in the vicinity of $E_{\rm F}$  the
numerical error in $E_{\rm F}$ position may be relatively large,
for this reason it makes sense to perform the calculations of the longitudinal 
${\sigma=\sigma_{xx}=\sigma_{yy}}$ and the spin Hall ${\sigma_{\rm  SH}=\sigma^{\uparrow}_{xy}-\sigma^{\downarrow}_{xy}}$ conductivities
as functions of energy within a finite energy window centered at $E_{\rm F}$. The results for $\sigma$, $\sigma_{\rm
  SH}$, and the spin Hall angle, $\sigma_{\rm SH}/\sigma$, are shown in
Figs.~\ref{fig2}\,c, d and e, respectively. Since the disorder-induced  broadening in this energy
window is also small, we will compute our transport quantities by 
adding a  small imaginary constant ${\delta={\rm Im}E}$ to the energy $E$,
and then subsequently decreasing it.  Smaller $\delta$ value will result into
a stronger oscillation of all energy-dependent quantities and the converged
calculation will require more $k$-points. For example, in case of
${\delta=1}$~meV, the $k$-mesh density was increased up to $10^{6}$, and
in case of ${\delta=0.1}$~meV -- to $10^7$ points in the irreducible wedge of the Brillouin zone.
By setting the $\delta$ value completely to zero, the remaining broadening
would be plainly disorder-induced, however at ${E\approx E_{\rm F}}$
this effect becomes so small, that the well-converged calculation of residual
conductivities would require an enormous number of $k$-points. For this reason we limit our
analysis to the results for ${\delta=1}$ and 0.1~meV.

\begin{figure}
\centering
\includegraphics[width=1.0\linewidth]{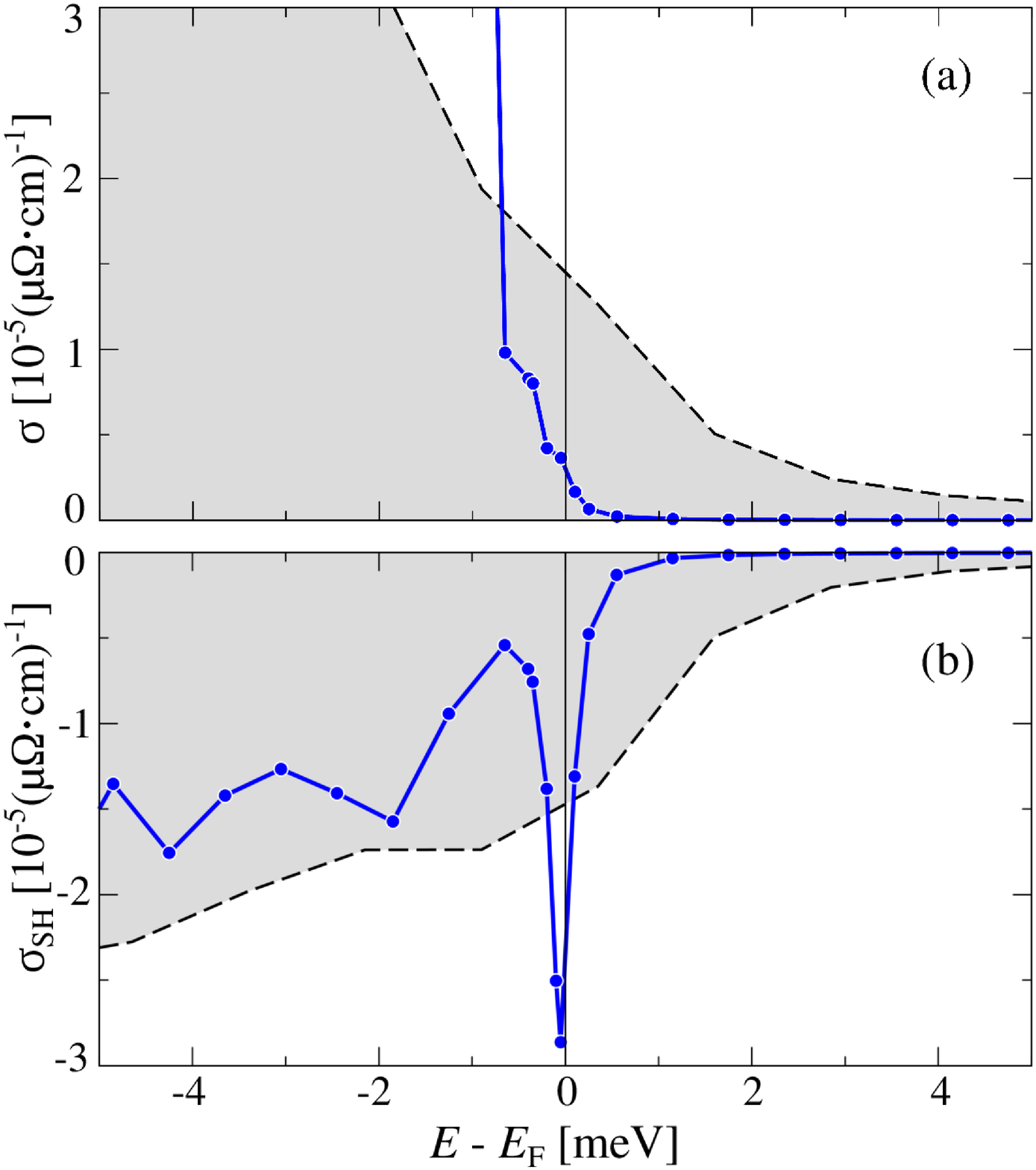}
\caption{(a)~Longitudinal conductivity $\sigma$ and (b) spin Hall conductivity $\sigma_{\rm SH}$, calculated
   as functions of energy $E$. Black dashed line corresponds to the imaginary energy offset  ${\delta=1}$~meV, thick blue line -- to ${\delta=0.1}$~meV.
}
\label{fig4}
\end{figure}

As follows from  Fig.~\ref{fig2}\,c, the mean amplitudes of
$\sigma$ and $\sigma_{\rm SH}$ as functions of energy within ${E<E_{\rm F}}$ strikingly
decrease by approaching $E_{\rm F}$. To a large extent this 
results from the DOS of the parabolic-shaped dispersion of the hole-like states. As follows from the BSF in
Fig.~\ref{fig2}\,a, they are Bloch-like and are perturbed by disorder
only slightly, thus remaining highly conductive and making the most substantial contribution to $\sigma$,
as well as to $\sigma_{\rm SH}$ (they consist mostly of $p$- and
$d$-symmetries, and thus ``feel'' the spin-orbit coupling) despite their
small DOS. By decreasing $\delta$  from 1~meV to 0.1~meV, the mean amplitude of $\sigma$ is amplified
 by a factor of about 10 times (since these Bloch-like states become even more Bloch-like), however its decrease towards $E_{\rm F}$
becomes more steep by leading to the same small values at
${E\approx E_{\rm F}}$. A more detailed plot in Fig.~\ref{fig4}\,a
reveals that for ${E\approx E_{\rm F}}$ this decrease is so steep that it produces even
smaller $\sigma$ values for ${\delta=0.1}$\,meV, thus the effects of $\delta$-broadening within
${E\ge E_{\rm F}}$ and ${E<E_{\rm F}}$ are opposite to each other.  
Indeed, for ${E\ge E_{\rm F}}$ the hole-like Bloch states are absent
and $\sigma$ is mostly defined by amplitudes of Lorentzian tails from the
hole-like states penetrating into this energy regime due to finite $\delta$ (and
disorder). 


Another strong impact on transport characteristics is produced by the
non-dispersive impurity-like states. Since they are localized,
they almost do not contribute to $\sigma$, but significantly contribute
to $\sigma_{\rm SH}$. This is especially evident for
${E=E_{\rm F}}$, where the hole-like Bloch-like states are
absent. By decreasing $\delta$, the Lorentzian tails get suppressed
together with their contributions to $\sigma$ and $\sigma_{\rm SH}$,
whereas the localized impurity states  get unscreened
and their contributions become, respectively, more pronounced. This is
clearly manifested by the sharp peak of  $\sigma_{\rm SH}$ at ${E\approx   E_{\rm F}}$ for ${\delta=0.1}$~meV (Fig.~\ref{fig4}\,b).

 Combination of these effects has interesting consequences for the
 spin Hall angle, ${\sigma_{\rm SH}/\sigma}$, which exhibits a divergent
 behavior at ${E\approx E_{\rm F}}$ (see Fig.~\ref{fig2}\,e) since
${\sigma\sim\delta + \sigma^{\rm imp}}$,  ${\sigma_{\rm
 SH}\rightarrow\sigma^{\rm imp}_{\rm SH}}$ and ${\sigma^{\rm imp}\ll\sigma^{\rm imp}_{\rm SH}}$, where ``imp'' refers to 
 the impurity. For ${\delta=0}$ the spin Hall angle will be maximized
being limited only by small Lorentzian broadening coming from
disorder. For  ${E<E_{\rm F}}$ the spin Hall angles are very small, even in
the presence of impurities (seen at about $-20$ and $-25$~meV, see
Figs.~\ref{fig2}\,a,\,b), since this situation is closer to that of a conventional metal,
 where $\sigma$ is large due to Bloch-like dominating contributions. For
 instance, as it was shown experimentally for Pt, it has quite large
 intrinsic  spin Hall conductivity ${\sigma_{\rm SH}\sim3\cdot10^{-4}\,(\mu\Omega\cdot\text{cm})^{-1}}$, but also large
 ${\sigma\sim4\cdot10^{-2}~(\mu\Omega\cdot\text{cm})^{-1}}$ (bounded by
  remaining impurities and finite temperature effects), which leads to the spin Hall
 angle of about $\sim 10^{-2}$~\cite{IVH+15}. Another set of the impurity
 states, seen as the sequence of many similar  stripes confined within a range of
 about 15~meV right below $E_{\rm F}$, does not
 contribute to $\sigma_{\rm SH}$ due to their $s$-symmetry.

The contribution of the conical states is difficult to distinguish: 
in the regime ${E>E_{\rm F}}$, where only conical
states exist, $\sigma$ strongly scales down with $\delta$, by indicating
that $\sigma$ is dominated by Lorentzian tails stemming from the 
parabolic states. On the other hand, by using the analogy with a charged particle moving
in a magnetic field (which includes the effect of the spin-orbit coupling), the relaxation time approximation ${\sigma_{xx}\sim\frac{ne^2}{\tau m}(\omega^2-\frac{1}{\tau^2})^{-1}}$, ${\omega\sim
  B_{\rm so}/m}$, $m$ - is the effective electron mass, shows that  $\sigma\rightarrow0$ for $m\rightarrow0$, i.e., for massless
particles. At the same time, the off-diagonal components ${\sigma_{xy}\sim\frac{ne^2\omega}{m}(\omega^2-\frac{1}{\tau^2})^{-1}\rightarrow\text{const}}$, which does not depend on the scattering time $\tau$, i.e. the
massless electrons will propagate along the equipotential lines with 
constant velocity which depends  on the spin-orbit coupling but does not
depend on the disorder scattering rate. However even this possible contribution
appears to be rather small, compared to the others. Indeed, despite that the conical bands look very much Bloch-like,
it is very difficult to distinguish even their DOS. It is not only
smaller than the DOS of the impurities, but it is also smaller than the DOS of the parabolic states for ${E<E_{\rm F}}$,
since in the vicinity of the $E_{\rm F}$:  ${n_{\rm par}\sim\sqrt{E-E_{\rm F}}}>{n_{\rm conical}\sim(E-E_{\rm F})^2}$. 

\section{Summary and Outlook}

Our detailed electronic structure calculations  for the  LaPt$_{0.5}$Pd$_{0.5}$Bi alloy suggest that the topological borderline half-Heusler
bulk systems with chemical disorder will necessarily exhibit 
impurity-like resonant states, with one of them located at the top of the valence band, 
which will coincide with the Dirac point at $E_{\rm F}$ due to
statistical restoring of the cubic symmetry. Such electronic
structure leads to a strong sensitivity of the electronic transport
characteristics with respect to the position of $E_{\rm F}$, which
might easily fluctuate in experiments. According to the presented first-principles calculations based
on the Kubo formalism, once the Fermi energy sticks to the impurity
level, it will result in a strong suppression of the
longitudinal conductivity $\sigma$, but will retain the off-diagonal
spin Hall
components $\sigma_{\rm SH}$, thus leading to divergent spin Hall
angles $\sigma_{\rm SH}/\sigma$. As we have seen, the suppression of $\sigma$ strongly
depends on the amplitude of the Lorentzian tails from the parabolic
Bloch-like states below the Fermi energy, which efficiently screen the impurity
effect. For this reason, the divergent behavior of the spin Hall angle
seems to be an essentially low-temperature feature when the additional broadening of
the Bloch-like states due to phonons is sufficiently weak. The
conical-dispersive Dirac bands appear to remain Bloch-like in agreement
with arguments in Refs.~\cite{NSPS14,SBB+15} and do not exhibit any
impurity-like effects, except the regime ${E\le E_{\rm F}}$ where they are
perturbed by the impurities connected with the parabolic bands. At the
same time, despite of their proximity to the Bloch states, their
own contribution to the conductivity components appear to be negligibly
small compared to the effects from the  Lorentzian tails
and the resonant impurities.  We stress, that the topological properties play here an
important, though an auxiliary role. In the nontrivial phase the system
would exhibit the impurity level at $E_{\rm F}$ as well, however due to
the proximity of the parabolic like bands (not only hole-like, but also
electron-like ones) the Lorentzian screening of the impurity will be too
strong. On the other hand, in the trivial case, $E_{\rm F}$  would be placed
in the band gap, i.e. above the impurity level.

\begin{figure}
\centering
\includegraphics[width=0.9\linewidth]{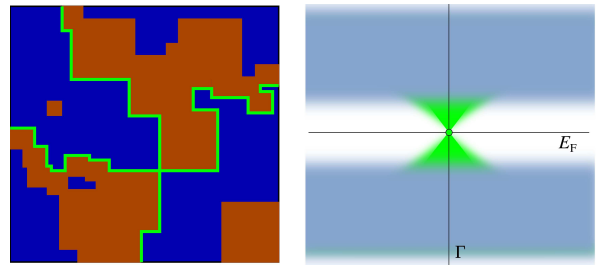}
\caption{The sketch of a topologically phase-separated disordered system (in two
  dimensions) at the percolation transition. Dark-blue and brown
  clusters represent topologically distinct phases. The system
  contains two distinct ``infinite'' clusters (connecting the opposite sites 
  of a square  situated infinitely far from   each other).  The green
  line marks the  infinite topological transition interface path
  along which the inversion of the energy levels occurs. The BSF of such
  system (corresponding already to a three-dimensional case, sketched on
  the right) will impose the Dirac state centered at $E_{\rm F}$
  (colored as green, the rest of the electronic states corresponds to the
  gray area); the broadening of the spectral density reduces by
  approaching the Dirac point.} 
\label{fig5}
\end{figure}

So far we have considered the properties of the homogeneously disordered
bulk systems, in which the bulk Dirac state is unstable with respect to
the stoichiometric variations. On the other hand, modeling of the
inhomogeneous disordered borderline systems may help to
overcome this problem. Let us consider the sketch in Fig.~\ref{fig5},
where the disordered system is interpreted as a topologically phase-separated 
system consisting of random clusters with trivial and
nontrivial order of their eigenstates. The topological inversion of states will inevitably
occur at the interface between topologically distinct clusters,
which represents a randomly curved two-dimensional manifold. 
Since the Dirac points distributed on this surface will possess a
two-dimensional nature, all considerations related to their
topological protection and the corresponding transport properties of the typical
Dirac surface states can be expected. The whole material will belong to the borderline topological state only, if
it contains at least two infinite connectible topologically distinct
clusters, since this guarantees an infinite and isotropic spread of the
Dirac-point surface, i.e. that it always connects the opposite sides
of a sample. The statistical conditions guaranteeing the existence of
such clusters are provided by percolation theory~\cite{Kir73}. For
example, in case of two-dimensional systems, the borderline state cannot be stabilized with respect to the
composition since it requires the relative amounts of both phases to be
precisely equal. For example, with respect to La(Pt$_{1-x}$Pd$_x$)Bi
alloys, the compositions with  ${x\neq0.5}$ will correspond to either trivial or topological
behavior of the whole system. In case of the three-dimensional  systems, the
percolation will persist within a large compositional range, 
${0.17<x<0.83}$, which greatly softens the restrictions for their experimental
synthesis.

\acknowledgments
The work was financially supported by the Deutsche Forschungsgemeinschaft Schwerpunksprogramm DFG-SPP~1666.


\end{document}